\documentclass{laynagenerictemplate}

\begin{document}

\maketitle \thispagestyle{empty}

\begin{abstract}
Meta-analysis is a well-established method for integrating results from several independent studies to estimate a common quantity of interest. However, meta-analysis is prone to selection bias, notably when particular studies are systematically excluded. This can lead to bias in estimating the quantity of interest. Motivated by a meta-analysis to estimate the rate of completed-suicide after bariatric surgery, where studies which reported no suicides were excluded, a novel zero-truncated count modelling approach was developed. This approach addresses heterogeneity, both observed and unobserved, through covariate and overdispersion modelling, respectively. Additionally, through the Horvitz-Thompson estimator, an approach is developed to estimate the number of excluded studies, a quantity of potential interest for researchers. Uncertainty quantification for both estimation of suicide rates and number of excluded studies is achieved through a parametric bootstrapping approach.\end{abstract}


\Addresses
\section{Introduction and background}
\label{sec:intro}

Meta-analysis \cite[e.g.][]{borenstein_etal_2021} is a statistical technique used to combine the results of several, independent studies to address a specific research question. For instance, the scenario considered in this paper is where meta-analysis is used to estimate the rate of a specific medical condition, using independent studies, each of whom was conducted to estimate the prevalence rate of the condition. The selection of studies to include in a meta-analysis is potentially subject to selection bias \cite[e.g.][Chapter 15]{kulinskaya_etal_2008}. Specifically, we consider the case where studies which did not observe any incidences of the medical condition are excluded from the meta-analysis leading to a misrepresentative estimated rate. Motivated by a meta-analysis to estimate the rate of completed suicide after bariatric surgery, we combine statistical methodology from the fields of meta-analysis and capture-recapture \cite[see][for more detail]{amstrup2005handbook, bohning2018capture, mccrea2014analysis} to develop new methodology to estimate the rate of completed suicide as well as the number of excluded studies.

Obesity is a growing global health issue, with 17.5\% of the world's population predicted to be classified as obese by 2030 \citep{lobstein2022world}. This increase in prevalence is likely to add pressure to healthcare systems, prompting policy-makers to seek effective treatments. One such treatment is bariatric surgery: medical procedures which work by at least one of the following mechanisms: (i) restricting food intake; (ii) decreasing nutrient absorption; or (iii) affecting cell signalling pathways leading to reduced appetite \citep{khwaja2010bariatric}. 

Any surgical procedure carries risk. After bariatric surgery, support and monitoring are required to ensure patients are able to adjust their lifestyle accordingly, especially if there are long term post-surgery complications. Significant lifestyle changes are necessary in order to ensure the best chance of success from the surgery, with many facing mental health issues as a result. 

Of particular interest \citep[see][and references therein]{peterhansel2013risk} is the risk of completed suicide post-bariatric surgery. A systematic review by \cite{peterhansel2013risk} provides meta-analytic data on 27 studies exploring the risk of completed suicide post-bariatric surgery. Table~\ref{tab:data} shows a snapshot of the data, with Table S1 in the Supplementary Material displaying the full dataset. The data include the number of completed suicides, the exposure (person-years), the country of origin and proportion of women of each study. A covariate for age is available, but due to being poor quality, it is not included in the main text. For completeness however, it is included in the table in the Supplementary Material with some analysis. From these data, a naive estimate (see Section~\ref{sec:metaanalysis}) of the rate of completed suicide is found to be 44.5 per 100,000 person-years. However, and crucially, data from studies with no observed completed suicides were excluded through only searching for studies with at least one completed suicide recorded. This results in an overestimation of the completed suicide rate and, ultimately, to misleading conclusions, impacting both medical professionals and patients.

Motivated by this case study, the paper develops new methodology for this problem by linking two research areas: model-based meta-analysis and capture-recapture population size estimation. We develop a zero-truncated count modelling approach to estimate the underlying suicide rate, which addresses observed heterogeneity through covariate modelling, and accounts for overdispersion caused by unobserved heterogeneity, where the latter describes the variation of rates of completed suicide across the studies which cannot be explained by random variation alone. This approach also allows estimation of the number of excluded studies (i.e. those with zero counts of completed suicide) for sub-populations defined by each unique set of observed covariates. Such estimates can be important for researchers investigating under-count occurrence. Uncertainty quantification for both estimation of suicide rates and number of excluded studies was achieved through a parametric bootstrapping approach.

\cite{peterhansel2013risk} considered a zero-truncated binomial model but did not consider covariate effects. We argue that a count model (e.g. Poisson or negative-binomial), with an offset to account for person-times at risk, is more appropriate for estimating underlying rates. An additional advantage of our modelling approach is it ability to estimate the number of excluded studies.

Section~\ref{sec:metaanalysis} briefly describes using meta-analysis to estimate rates. Section~\ref{sec:zerotrunc} introduces the zero-truncated count modelling approach and the estimation of the number of excluded (or missing) studies is addressed in Section~\ref{sec:missing}.

\begin{table*}[htbp]
\centering
\caption{An overview of the meta-analytic data from \cite{peterhansel2013risk}, numbered and ordered by decreasing size of person-years. The table includes the number of person-years, the proportion of women, the country of origin and the number of completed suicides for each study. The full dataset is available in Table~S1 in the Supplementary Material. \label{tab:data}}
{\small
\begin{tabular}{@{}lrrrrl@{}}
\toprule
Study & Person-& Proportion & Country & Number of \\
 & years & of women & of origin & completed suicides\\
  $i$& $e_i$ & $x_{i1}$ & $x_{i2}$&$y_i$\\
\hline
1. Adams 2007 & 77602 & 0.860 & USA & 21\\ 
2. Marceau 2007 & 10388 & 0.720 & Canada & 6\\ 
\vdots & \vdots & \vdots & \vdots & \vdots \\
26. Svenheden 1997 & 166 & 0.791 & Sweden & 1\\ 
27. Pekkarinen 1994 & 146 & 0.704 & Finland & 1\\
\bottomrule
\end{tabular}}
\end{table*}

\section{Meta-analysis} \label{sec:metaanalysis}

Briefly, meta-analysis is the methodology for integrating results from a number of independent studies focused on the same research question. Conventionally, there are two main elements, finding the summary statistics for each study, then combining these statistics into an overall weighted average. This can be beneficial, especially for rare outcomes, e.g. completed suicide, since the precision in estimating the prevalence rate can be increased when compared to any single study. 

In our case study, there are $n=27$ studies and we let $y_i$ and $e_i$ denote the observed number of completed suicides and person-years for the $i$th study, respectively, for $i=1,\dots,n$. Suppose $y_i$ is a realization of the random variable $Y_i$, for $i=1,\dots,n$, and it is assumed throughout that $Y_1,\dots,Y_n$ are independent. Lastly, let $\bx_i = (x_{i1},x_{i2})^T$ denote the covariates for the $i$th study, where $x_{i1}$ is the proportion of women and 
$$x_{i2} = \left\{ \begin{array}{ll}
1 & \mbox{if country of origin is USA;}\\
0  & \mbox{if otherwise,}
\end{array}\right.$$
for $i=1,\dots,n$.

If it is assumed that $Y_i \sim \mathrm{Poisson}\left(e_i \rho \right)$ where $\rho$ is the population suicide rate (the same for all sub-populations, irrespective of covariates), then the maximum likelihood estimator of $\rho$ is
\begin{equation}
\hat{\rho} = \frac{\sum_{i=1}^n Y_i}{\sum_{i=1}^n e_i}.
\label{eqn:miss1}
\end{equation}
This can also be seen \citep[e.g.][]{barendregt_etal_2013} as a weighted average of $\hat{\rho}_1, \dots,\hat{\rho}_n$, where $\hat{\rho}_i = Y_i/e_i$ is the $i$th study-specific estimator of $\rho$, and where the weights are the inverse variances of $\hat{\rho}_1, \dots,\hat{\rho}_n$ under the Poisson assumption.

An alternative estimator \citep[e.g.][]{cooper2019handbook, egger2008systematic, stangl2000meta, borenstein_etal_2021} can be formed by operating on the log scale, i.e. an estimator of $\log(\rho)$ is given by a weighted average of $\log (\hat{\rho}_1), \dots, \log (\hat{\rho}_n)$. The weights are given by the inverse variances of $\log (\hat{\rho}_1), \dots, \log (\hat{\rho}_n)$ under the Poisson assumption, and approximated by using the delta method. In this case
\begin{equation}
\hat{\rho} = \exp \left[\frac{\sum_{i=1}^n Y_i \log \left(Y_i/e_i \right)}{\sum_{i=1}^n Y_i}\right].
\label{eqn:miss2}
\end{equation}
For the data in Table~\ref{tab:data} in the supplementary material, the corresponding two estimates from (\ref{eqn:miss1}) and (\ref{eqn:miss2}) are 44.5 and 60.0 per 100,000 person-years, respectively.

To account for potential covariate effects, a generalized linear model (\citealt{mccullagh_nelder_1989}) can be fitted under a Poisson response distribution, a log link function and $\log (e_1),\dots,\log (e_n)$ as offsets, i.e. $Y_i \sim \mathrm{Poisson}\left(e_i \rho_i\right)$, where $\rho_i = \exp \left(\eta_i \right)$ and $\eta_i = \mathbf{h}(\mathbf{x}_i)^T \boldsymbol{\beta}$ is the linear predictor. The regression function $\bh(\cdot)$ can be used to implement different effects of the covariates. We consider five different linear predictors as shown in Table~\ref{tab:lp}.

\begin{table*}[htbp]
\centering
    \caption{Linear predictors considered in this paper. }
    \label{tab:lp}
    {\small
    \begin{tabular}{@{}llllllll@{}}
    \toprule
Linear & Proportion & Country & Interaction & $\bh(\bx)$ \\ 
predictor & of women, $x_1$ & of origin, $x_2$ & $x_1x_2$ & \\ \hline 
1 & No & No & No & $\bh_1(\bx) = 1$\\
2 & Yes & No & No & $\bh_2(\bx) = (1,x_1)^T$\\
3 & No & Yes & No & $\bh_3(\bx) = (1,x_2)^T$ \\
4 & Yes & Yes & No & $\bh_4(\bx) = (1,x_1,x_2)^T$ \\
5 & Yes & Yes & Yes& $\bh_5(\bx) = (1,x_1,x_2,x_1x_2)^T$ \\ \bottomrule
    \end{tabular}}
\end{table*}

One study, 24. Smith 2004, did not report a proportion of women. An imputation model was fitted to impute this missing value. The linear regression imputation model has proportion of women as a response and considers country of origin, person-years and number of complete suicides (and all two-way interactions) as covariates. The model minimizing the Bayesian information criterion (BIC; e.g. \citealt{davison_2003}, page 152) has main effects for person-years and country of origin, and their interaction. From this model, the imputed value of proportion of women was $x_{24,1} = 0.823$. Furthermore, study 21. Kral 1993, reported a country of origin as USA/Sweden. This was changed to solely USA. Both of these changes are indicated by italics in Table~S1 in the Supplementary Material.

For the linear predictors in Table~\ref{tab:lp}, the model that minimizes the BIC statistic is linear predictor 3, seemingly indicating that the suicide rate is different for sub-populations in USA to other countries. Furthermore, the estimated suicide rate for sub-populations in USA is 34.6 and in other countries is 68.1 (both per 100,000 person-years).

If there is only one study, there is a clear rationale for the Poisson model with an offset and the parameter estimate corresponds to the observed rate. The Poisson model implies that the variance of $Y_i$ is equal to the mean which might hold for one study but cannot be tested.  However, if there are many studies, the rate of completed suicide may vary across the studies leading to unobserved heterogeneity and consequently, overdispersion \citep[e.g.][Section 10.6]{davison_2003}. An alternative model is to assume a negative-binomial distribution where the variance of $Y_i$ is $e_i \rho_i + e_i^2 \rho_i^2/\alpha$, where $\alpha > 0$ is an unknown dispersion parameter and if $\alpha$ becomes large, the negative-binomial model approximates the Poisson. We fitted five models under the same five linear predictors as in Table~\ref{tab:lp} but under a negative-binomial distribution. However, the BIC values did not indicate an improved fit, compared to the Poisson distribution, when balanced against the extra complexity of the model (one extra parameter: $\alpha$).

Uncertainty quantification can be achieved using asymptotic results or via bootstrapping. However, the models fitted in this section ignore the excluded (or missing) studies with zero counts of completed suicides. This can lead to overestimation of the suicide rate. In the next section, we develop a zero-truncated count modelling approach to take account of the excluded studies. Additionally, this methodology also has the capability to estimate the number of excluded studies (see Section~\ref{sec:missing}).

\section{Zero-truncated modelling} \label{sec:zerotrunc}

\subsection{Model development}\label{sec:moddev}

In this section, we develop a zero-truncated count modelling approach. We assume that $Y_i$ is distributed according to a zero-truncated count distribution with probability function 
$$p(y ; \mu_i, \alpha)^+ = \frac{p(y ; \mu_i, \alpha)}{1 - p(0 ; \mu_i, \alpha)},$$
for $y = 1, 2, \dots$, where $p(\cdot; \mu_i,\alpha)$ is the probability function of a count distribution with support given by the non-negative integers and 
mean $\mu_i$, possibly depending on a dispersion parameter $\alpha$. Two examples of such a distribution, considered in this paper, are the Poisson and negative-binomial. The Poisson distribution has probability function $p_P(y; \mu) = \exp(-\mu)\mu^y/y!$, (i.e. not depending on $\alpha$) and the negative-binomial has 
$$p_{NB}(y; \mu, \alpha) = \frac{\Gamma(\alpha+y)}{y!\Gamma(\alpha)} \left[\frac{\alpha}{(\alpha+\mu)}\right]^{\alpha} \left[\frac{\mu}{(\alpha+\mu)}\right]^{y},$$
where $\Gamma(\cdot)$ is the Gamma function.

\begin{table*}[ht]
\centering
    \caption{Values of the maximized log-likelihood, number of parameters,  AIC statistic, BIC statistic and BIC weights for models under consideration. BIC weights are only given for the Poisson and negative-binomial models since binomial models are not suitable for estimating rates.  \label{tab:bic}}
    {\footnotesize
    \begin{tabular}{@{}lrrrrrr@{}}
    \toprule
   Distribution & Linear predictor & Maximized & Number of & AIC & BIC & BIC\\
     & & log-likelihood & parameters & & &  weights \\ \hline
     & 5 & -22.7 & 4 & 53.4 &  58.6 & 0.0097\\
     & 4 & -23.0 & 3 &  52.0 & 55.9 & 0.0362\\
Poisson  & 3 &  -23.0 & 2 &  50.1 & 52.6 & 0.1863\\
     & 2 &  -23.4 & 2 &  50.9 & 53.4 & 0.1251\\
     & 1 & -23.7 & 1 &  49.5 & 50.7 & 0.4813\\ \hline     
      & 5 & -22.7 & 5 &  55.4 & 61.9 & 0.0019\\
      & 4 & -23.0 & 4 &  54.0 & 59.2 & 0.0070\\
Negative-binomial & 3 & -23.0 & 3 &  52.1 & 55.9 & 0.0359\\
     & 2 & -23.4 & 3 &  52.9 & 56.7 & 0.0241\\
     & 1 & -23.7 & 2 & 51.5 & 54.0 & 0.0926\\ \hline
     & 5 & -22.7 & 4 &  53.4 & 58.6 & -\\
     & 4 & -23.0 & 3 &  52.0 & 55.9 & -\\
Binomial  & 3 & -23.0 & 2 &  50.1 & 52.6 & -\\
    & 2 & -23.4 & 2 &  50.9 & 53.4 & -\\
     & 1 & -23.7 & 1&  49.5 & 50.7 & - \\ \bottomrule
    \end{tabular}}
\end{table*}

We assume $\mu_i = e_i \rho_i = e_i \exp \left[ \bh(\mathbf{x}_i)^T \boldsymbol{\beta} \right]$ and consider the five linear predictors given in Table~\ref{tab:lp} with both the zero-truncated Poisson and negative-binomial distributions. Table~\ref{tab:bic} shows the values of  AIC, BIC, maximized log-likelihood and number of parameters of the models considered. There is negligible improvement (in terms of maximized log-likelihood) of the negative-binomial models compared to Poisson. This can be observed from the dispersion parameter, $\alpha$, being estimated to take large values under each of the negative-binomial models. For example, under the intercept-only (linear predictor 1) zero-truncated negative-binomial model, $\hat{\alpha} = 1.9\times 10^5$. This indicates that the response variance is very close to the mean, indicative of the Poisson model being adequate. The same conclusion was drawn from the non-zero-truncated models fitted in Section~\ref{sec:metaanalysis}. However, unlike in Section~\ref{sec:metaanalysis}, there appears to be no significant covariate effect on the suicide rate, with the intercept-only zero-truncated Poisson model minimizing both the AIC and the BIC statistic, and hence being the preferred model. It should be noted that in general, if there is covariate information available, it is beneficial to include this information in the modelling and estimation as the additional information can aid in improving the accuracy of the results. Whilst in this case, the covariate information is not significant, this would not have been known without first investigating the models with covariate information included. In this way, we can gain more trust in the validity of the intercept-only model.

Both the Poisson and negative-binomial models looked at are identifiable, with the best fitting model always selected as the Poisson with an offset. This model is used for the inference, and given that the model is valid for the observed data, it is plausible to assume that the model is also valid for the unobserved data to estimate $p(0;\mu_i, \alpha)$, though this assumption cannot be tested.

\begin{sloppypar}For linear predictor $j=1,\dots,5$ and response distribution $D \in \big\{ \text{Poisson (P)}, \text{ negative-binomial (NB)} \big\}$, the maximum likelihood estimate of the suicide rate for a sub-population with covariates $\bx$ is given by
$$\hat{\rho}_{\bx} = \exp \left[ \mathbf{h}_j(\mathbf{x})^T\hat{\boldsymbol{\beta}}^{(D)}_j \right],$$
where $\hat{\boldsymbol{\beta}}_j^{(D)}$ are the maximum likelihood estimates of the parameters under linear predictor $j=1,\dots,5$ and distribution $D$. The intercept-only zero-truncated Poisson model leads to a maximum likelihood estimate of the suicide rate which is constant for all covariates $\bx$ and is given by $\hat{\rho} = 31.8$ (per 100,000 person-years). \end{sloppypar}
Within capture-recapture studies, the total number of individuals is always unknown, so it is difficult to evaluate the accuracy of the results. However, in these studies, the fact that the total number is hidden provides the motivation and need for the estimation methods. In this case, the resulting zero-truncated intercept-only Poisson model fits the observed data well, so if it is assumed that this well-fitting model extends to the unobserved zero counts, then a valid estimate of the missing number of studies can be achieved. The estimate of 31.8 completed suicides per 100,000 person-years is supported by the Expectation-Maximization (EM) algorithm, given that it results in the exact same value for the maximum likelihood estimate used to find the same number of missing studies (see Supplementary material for EM algorithm).

\cite{peterhansel2013risk} considered a zero-truncated binomial distribution, where the number of trials is the person-years, $e_i$, and the ``success" probability is the suicide rate, $\rho$, constant for all sub-populations. We extend this approach by letting the success probability be $\rho_i = \{ 1 + \exp [ -\bh(\mathbf{x}_i)^T \boldsymbol{\beta} ] \}^{-1}$ with the same linear predictors given in Table~\ref{tab:lp}. The resulting maximized log-likelihood, number of parameters and BIC and AIC statistics are shown in Table~\ref{tab:bic}. The preferred zero-truncated binomial model is the model with only an intercept and this model leads to an estimated suicide rate identical (to 1 decimal place) to the estimated rate under the intercept-only zero-truncated Poisson model (along with very similar maximized log-likelihood, AIC and BIC). This is because for low suicide rates, the Poisson and binomial models approximately coincide. It can be seen in the Supplementary Material that the non-zero-truncated Poisson and Binomial maximum likelihood estimates are equal, however, for the zero-truncated Poisson and binomial distributions, it is coincidence. Note that \cite{peterhansel2013risk} reported an estimate of 41 (per 100,000 person-years) from the same model which we are unable to reproduce.

\subsection{Assessing model adequacy}

To consider the adequacy of the preferred intercept-only zero-truncated Poisson model, we compare the observed and fitted frequencies of completed suicides. For $y=1,2,\dots, m$, where $m$ is the largest observed count, let $r_y = \sum_{i=1}^n I(y_i = y)$ be the number of studies with $y$ completed suicides, where $I(A)$ denotes the indicator function for event $A$, and let the fitted frequencies be given by
$$\hat{r}_y =\sum_{i=1}^n p_P\left(y; e_i \exp \left[ \bh_1(\bx_i)^T \hat{\boldsymbol{\beta}}_1^{(P)}\right] \right)^+ =\sum_{i=1}^n p_P\left(y; e_i \exp \left[\hat{\beta}_1^{(P)} \right] \right)^+,$$
where $\hat{\boldsymbol{\beta}}_1^{(P)} = \hat{\beta}_1^{(P)} = -8.055$ is the estimated intercept parameter under the intercept-only zero-truncated Poisson model.

\begin{table*}[htbp]
    \centering
    \caption{Frequency distribution for observed and fitted count of completed suicide, with the frequencies of more than or equal to 4 counts grouped into one category. \label{tab:fittedfreq}}
    {\small
    \begin{tabular}{@{}lrrrrr@{}}
    \toprule
        Count of Completed Suicide & 0 & 1 & 2 & 3 & 4+
        \\
         \hline
         Observed Frequency, $r_y$ & - & 18 & 3 & 3 & 3\\
         Fitted Frequency, $\hat{r}_y$ & - & 18.3 & 4.5 & 1.7 & 2.5\\
         \bottomrule
    \end{tabular}}
\end{table*}
Table~\ref{tab:fittedfreq} shows the observed and fitted frequencies where numbers of suicides greater or equal to four have been combined. Goodness-of-fit can be formally assessed using a $\chi^2$ test, where the test statistic is $\chi^2 = \sum_{y=1}^4 \left(r_y - \hat{r}_y\right)^2/\hat{r}_y$. Under the null hypothesis that the model is adequate, the distribution of $\chi^2$ is $\chi^2_2$. The observed value of $\chi^2$ is 1.59 leading to a $p$-value of 0.45 and there being no evidence of model inadequacy.

\subsection{Uncertainty quantification} \label{sec:boot}

To quantify the uncertainty in the estimation of the suicide rate under the Poisson intercept-only model, the endpoints of an approximate Wald-type $(1-\alpha)100$\% confidence interval for the suicide rate $\rho$ are obtained as follows
$$\exp \left[ \hat{\beta} \pm z_{1-\alpha/2} \mathrm{se}\left(\hat{\beta}\right) \right],$$
where $z_{1-\alpha/2}$ is the $(1-\alpha/2)$th quantile of the standard normal distribution and $\mathrm{se}\left(\hat{\beta}\right)$ is the standard error of $\hat{\beta}$. Under this approach, a 95\% confidence interval for $\rho$ is $(23.3, 43.2)$ per 100,000 person-years.

\begin{sloppypar}However, using the detection method of \citet{yee_2022}, the intercept in the Poisson intercept-only model suffers from the Hauck-Donner effect \citep{hauck_donner_1977}. This is where the Wald test statistic fails to increase monotonically as a function of its distance from $\beta = 0$. This can render Wald-type inference unreliable. Furthermore, the Wald approach does not take account of model uncertainty.\end{sloppypar}

Instead, we use bootstrapping. For this work, the parametric bootstrap algorithm is used in favour of the non-parametric. This is because whilst there is little difference between using a parametric and a non-parametric bootstrap algorithm for uncertainty quantification for the rate, when quantifying the uncertainty for the estimated missing number of studies, there is high correlation between the covariate combinations leading to inflated upper limits when the non-parametric approach is used (see Supplementary Material for non-parametric results). This procedure samples $n$ counts from the fitted model chosen with probability equal to its BIC weight (see Table~\ref{tab:bic}), given that BIC weights can be interpreted as conditional probabilities for use in model selection \citep{wagenmakers2004aic}. Whilst for the model selection there is no difference between the AIC and BIC statistics in terms of ordering the preference of the models, for bootstrapping, BIC weights are favoured over AIC weights given that the BIC focuses more on selecting the \textit{true} model \citep{chakrabarti2011aic}. We then fit each of the ten competing models given by the zero-truncated Poisson and negative-binomial models under the five different linear predictors given in Table~\ref{tab:lp} using the resampled counts and observed explanatory variables. The model with the lowest BIC is selected, and the estimated suicide rate for sub-populations of interest are recorded. This procedure is repeated $B$ times, where we set $B=25,000$  to exclude any possible random error induced by the Monte Carlo method and ensure that the number of iterations was large enough such that any additional iterations would not significantly alter the results.

\begin{figure}[htbp]
    \centering
    \includegraphics[width=0.65\linewidth]{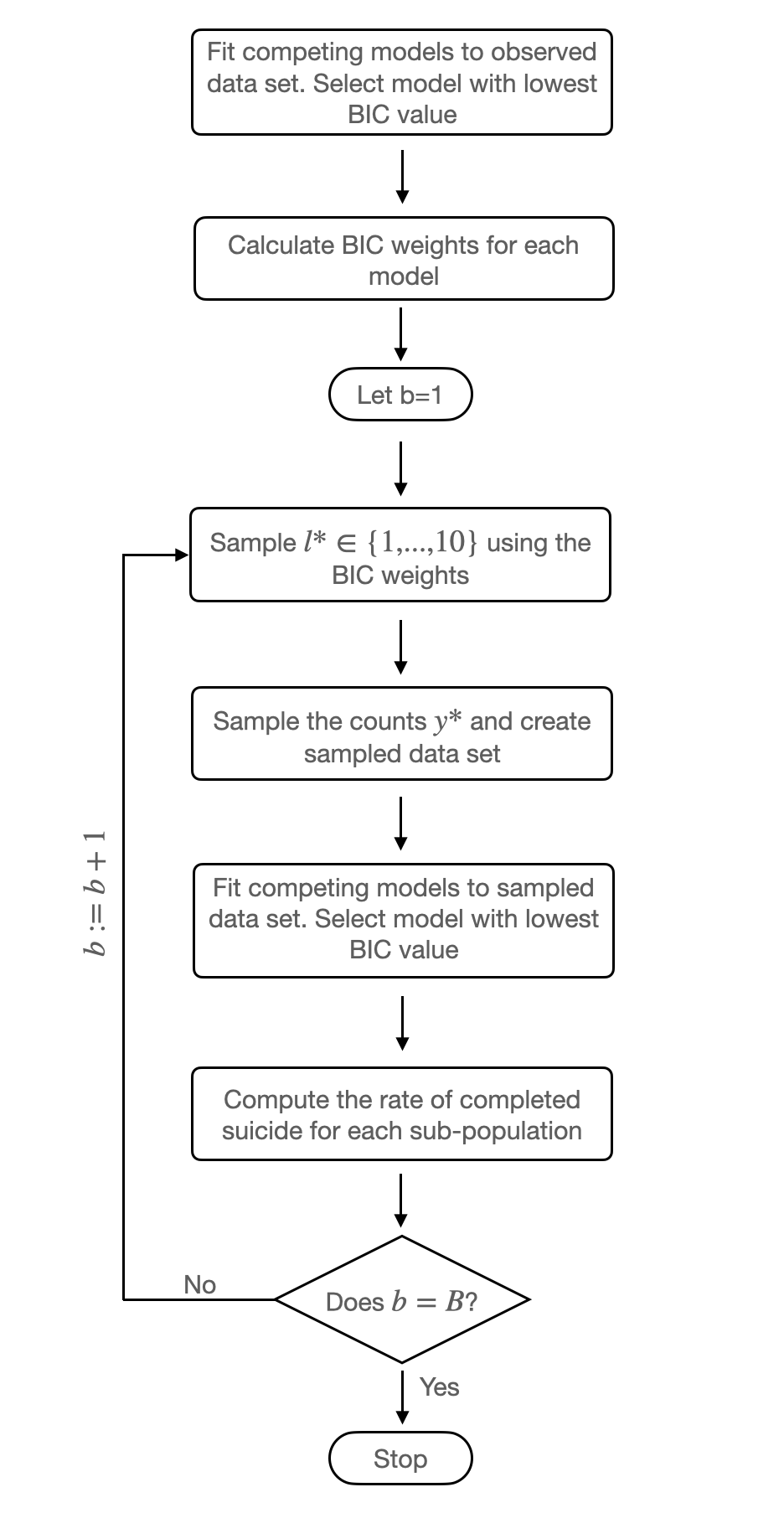}
    \caption{Flow chart for the parametric bootstrap algorithm}
    \label{fig:bootflow}
\end{figure}

Since a model with covariates may be selected for the resampled data, the estimated suicide rates may be different for different sub-populations. This means bootstrap confidence intervals for the rates may be different even though the intercept-only Poisson model (from Section~\ref{sec:moddev}) estimate these to be constant for all sub-populations. We choose six different sub-populations defined by the combination of country of origin being USA or other (not USA), and proportion of women being 0.75, 0.8 and 0.85. These three proportions were chosen as they (approximately) corresponded to the three quartiles of the observed proportion of women. For completeness, the parametric bootstrap algorithm is given below with a simplified version of the parametric bootstrap algorithm given in Figure~\ref{fig:bootflow}.

Let $\bar{\bx}_1 = (0.75, 1)^T$, $\bar{\bx}_2 = (0.75, 0)^T$, $\bar{\bx}_3 = (0.8, 1)^T$, $\bar{\bx}_4 = (0.8,0)^T$, $\bar{\bx}_5 = (0.85,1)^T$, and $\bar{\bx}_6 = (0.85,0)^T$ be the covariates for the six sub-populations.

\begin{enumerate}
    \item 
    Fit the ten competing models to the observed data and record the BIC value for each model. The best candidate model is the one with the lowest BIC value.
    \item
    For $l=1,...,10$, calculate the BIC weights for each model as
    \begin{equation*}
        w_l=\dfrac{\exp\left[-\frac{1}{2}\Delta_l(BIC)\right]}{\sum_{k=1}^K\exp\left[-\frac{1}{2}\Delta_k(BIC)\right]},
    \end{equation*}
    where $\Delta_l(BIC)=BIC_l-\text{min}(BIC)$ is the difference between the BIC value for each model and the BIC of the best candidate model for $l=1,...,10$.
    \item
    Let $b=1$.
    \item 
    \begin{sloppypar}Sample the value $l^*$ from $\{1,2,...,10\}$ with respective probabilities $\{w_1, w_2, ..., w_{10}\}$. Let $(j^*, D^*)$ be the pair of linear predictor and distribution chosen by
    $l^*$, which takes values $(l^*, P)$ if $l^*\leq 5$ and $(l^*-5, NB)$ otherwise.\end{sloppypar}
    \item 
    \begin{sloppypar}
    For $i=1,\dots,n$ sample $y^*_i$ from the distribution with probability function $p_{D^*}^+\left(\cdot ; e_i \exp \left[ \bh_{j^*}(\bx_i)^T \hat{\boldsymbol{\beta}}_{j^*}^{(D^*)}\right],\hat{\alpha}_{j^*}^{(D^*)}\right)$, where $\hat{\alpha}_{j^*}^{D^*}$ is the estimate of the dispersion parameter if $D^*=(NB)$. Create a new dataset $\left\{(e_1,\bx_1, y^*_1), \dots, (e_n, \bx_n, y^*_n) \right\}$.
    \end{sloppypar}
    \item
    Fit the ten competing models. Let $(\tilde{j}, \tilde{D})$ be the pair of linear predictor and distribution that minimizes the BIC. Let $\hat{\boldsymbol{\beta}}_{\tilde{j}}^{(\tilde{D})}$ be the maximum likelihood estimates of $\boldsymbol{\beta}_j$ for the selected model. 
    
    Calculate the estimated suicide rates for the six sub-populations using
    $$\rho^*_{bk} = \exp \left[ \bh_{\tilde{j}}\left(\bar{\bx}_k\right)^T \hat{\boldsymbol{\beta}}_{\tilde{j}}^{(\tilde{D})}\right],$$
    for $k=1,\dots,6$.
    \item
    If $b=B$, stop. Otherwise, return to Step 4 and let $b:=b+1$.
\end{enumerate}

\begin{table*}[htbp]
    \caption{95\% percentile bootstrap confidence intervals for the suicide rates (per 100,000 person years) for the six sub-populations. \label{tab:rateci}}
    \centering
    \begin{tabular}{@{}lccc@{}}
    \toprule
      & \multicolumn{3}{c}{Proportion of women} \\ \cline{2-4}
      Country of origin & 0.75 & 0.80 & 0.85 \\ \hline
      USA & $(17.9, 42.6)$ & $(19.1, 42.6)$ & $(19.2, 42.5)$ \\ 
      Others & $(15.7, 56.4)$ & $(16.3, 59.2)$ & $(17.1, 59.2)$  \\ \bottomrule
    \end{tabular}
\end{table*}

Table~\ref{tab:rateci} shows the 95\% percentile bootstrap confidence intervals for the suicide rate in each of the six sub-populations of interest. Typically, the confidence intervals are approximately centred at the common estimate of 31.8 (per 100,000 person years). Uncertainty can be deduced by the length of the confidence intervals. This indicates that, for the USA, uncertainty decreases as the proportion of women increases, whereas, for other countries, uncertainty remains fairly constant as proportion of women increases. This is plausible since, for the USA, 90\% of the studies have proportion of women greater than both 0.75 and 0.80. These values for other countries are 65\% and 41\%, respectively. The larger amount of data available leads to increased precision.

\section{Estimating the number of excluded studies} \label{sec:missing}

In Section~\ref{sec:zerotrunc}, methods to account for the exclusion of studies with zero counts of completed suicides were developed. Of interest is the estimation of the number of excluded studies (possibly stratified by sub-populations). We address this estimation in this section.

\subsection{Horvitz-Thompson estimation} \label{sec:ht}

For linear predictor $j= 1,\dots,5$, the total number of studies with the same covariates, $\bx_i$, and person-years, $e_i$, as study $i=1,\dots,n$, is estimated by the Horvitz-Thompson estimator (\citealt{horvitz1952generalization}, see also, e.g. \citealt[][Chapter 3]{mccrea2014analysis} or \citealt[][Chapter 11]{borchers2002estimating}), given by
$$\hat{N}_i = \frac{1}{1 - p_P\left(0; e_i \exp \left[ \bh_j(\bx_i)^T \hat{\boldsymbol{\beta}}_j^{(P)} \right] \right)},$$
for a Poisson model \citep[see][for more detail]{van2003point} and by
$$\hat{N}_i = \frac{1}{1 - p_{NB}\left(0; e_i \exp \left[ \bh_j(\bx_i)^T \hat{\boldsymbol{\beta}}_j^{(NB)} \right], \hat{\alpha} \right)},$$
for a negative-binomial model \citep[see][for more detail]{cruyff2008point}.

The number of excluded studies is then estimated by $\hat{M}_i = \hat{N}_i-1$, where the minus one corresponds to the non-excluded $i$th study. Under the preferred intercept-only Poisson model identified in Section~\ref{sec:zerotrunc}, Table~\ref{tab:missing} shows the estimated number of excluded studies for eight sub-populations defined by the combination of the country of origin being USA or other (not USA), and proportion of women being in the intervals $[0,0.75)$, $[0.75,0.80)$, $[0.8, 0.85)$ and $[0.85,1]$. These intervals are defined by the three approximate sample quartiles of the observed proportion of women.

\begin{table*}
    \caption{Estimates for the number of missing studies with corresponding 95\% percentile bootstrap confidence intervals for the eight sub-populations of interest, subtotals and totals. Also shown are the number of observed studies in [square brackets].\label{tab:missing}}
    \centering
    {\small
    \begin{tabular}{@{}lrrrrr@{}}
    \toprule
     & \multicolumn{4}{c}{Proportion of women} &  \\ \cline{2-5}
       Country of origin & $[0,0.75)$ & $[0.75,0.80)$ & $[0.8, 0.85)$ & $[0.85,1]$ &Total \\ \hline
  \multirow{2}{*}{USA} & 0 [1] & 0 [0] & 22 [6] & 8 [3] & 30 [10]\\
      & (0, 1) & (0, 0) & (16, 36) & (6, 13) & (22, 50) \\
     \multirow{2}{*}{Others} & 42 [6] & 23 [4] & 7 [4] & 5 [3] & 77 [17] \\    
      & (23, 115) & (12, 46) & (3, 15) & (2, 12) & (41, 206) \\\hline
     \multirow{2}{*}{Total} & 42 [7] & 23 [4] & 29 [10] & 13 [6] & 107 [27]\\ 
      & (23, 136) & (12, 46) & (21, 49) & (9, 23) & (74, 237) \\ \bottomrule
    \end{tabular}}
\end{table*}

The estimates in Table~\ref{tab:missing} show that it is unlikely there are many excluded studies originating in the USA with a proportion of women participants less than 80\%. Conversely, for studies not originating in the USA, it is estimated that most excluded studies have a proportion of women participants less than 80\%. 

\subsection{Uncertainty quantification} \label{sec:boot2}

The bootstrap algorithm can be used to quantify the uncertainty from using capture-recapture estimators \citep[see][]{buckland1991quantifying, zwane2003implementing, efron1981censored} in the estimation of the number of excluded studies. In this section, we use a parametric bootstrap similar to Section~\ref{sec:boot}, accounting for model uncertainty through including each of the models under consideration to reduce the risk of underestimating the variance. To quantify the uncertainty in the estimation of the number of excluded studies using each of the ten models under consideration, Step 6 of the bootstrapping algorithm presented in Section~\ref{sec:boot} is modified to be as follows.

\begin{enumerate}
    \item[6.]
    Fit the ten competing models. Let $(\tilde{j}, \tilde{D})$ be the pair of linear predictor and distribution that minimizes the BIC. Let $\hat{\boldsymbol{\beta}}_{\tilde{j}}^{(\tilde{D})}$ be the maximum likelihood estimates of $\boldsymbol{\beta}_j$ for the selected model. If $\tilde{D} = (NB)$, let $\hat{\alpha}$ be the estimate of the dispersion parameter.

    For $i=1,\dots,n$, calculate the estimated number of excluded studies with the same covariates and person-years as study $i$, using $\hat{M}^*_{bi} = \hat{N}^*_{bi} - 1$, where
    \begin{equation*}
        \hat{N}_{bi}^* = 
        \begin{cases}
            \dfrac{1}{1 - p_P\left(0; e_i \exp \left[ \bh_{\tilde{j}}(\bx_i)^T \hat{\boldsymbol{\beta}}_{\tilde{j}}^{(P)} \right] \right)} & \text{if }\tilde{D} = P, \\[18pt] 
            \dfrac{1}{1 - p_{NB}\left(0; e_i \exp \left[ \bh_{\tilde{j}}(\bx_i)^T \hat{\boldsymbol{\beta}}_{\tilde{j}}^{(NB)} \right], \hat{\alpha} \right)} & \text{if }\tilde{D} = NB.
        \end{cases}
    \end{equation*}
    The estimated total number of studies is calculated as 
    \begin{equation*}
        \hat{N}_b^*=\sum_{i=1}^n \hat{N}_{bi}^*,
    \end{equation*}
    with corresponding estimated total number of missing studies $\hat{M}_b^*=\hat{N}_b^*-n$.
\end{enumerate}

Note that some covariate combinations can occur repeatedly (see Table~\ref{tab:missing}). Therefore, the total number of studies for a specific sub-population (a specific covariate combination), are the sum of $\hat{N}_{bi}^*$ over all $i$ with this specific covariate combination.

Using the percentile method, the $95\%$ percentile confidence interval for the total number of studies is $(101, 264)$, with corresponding interval for the number of missing studies $(74,237)$. These intervals are centred around the Horvitz-Thompson estimate of the total number of studies, 134, and the estimated number of missing studies in Table~\ref{tab:missing}, 107, respectively. Additionally, $95\%$ percentile confidence intervals for the sub-population size estimates, seen in Table~\ref{tab:missing}, are calculated and show that there is less uncertainty for estimates originating from the USA given that the total confidence intervals are narrower. The intervals also show that studies with higher proportions of women, typically have less uncertainty. These conclusions are plausible given that more observed studies originate from the USA than others in addition to the majority of observed studies having a proportion of women greater than the approximate median of 80\%. 

\section{Discussion}\label{sec: discus}

This paper proposes a novel zero-truncated count modelling approach to estimate a prevalence rate for meta-analytic data, taking account of excluded (or missing) studies with zero counts of the condition. This approach can incorporate covariate effects (with estimates of the prevalence rate stratified by covariates) and overdispersion. The approach has the ability to also estimate the number of excluded studies (again stratified by covariates). A bootstrapping procedure is developed to provide uncertainty quantification. In this paper, the approach is demonstrated by being used to estimate the prevalence of completed suicide post-bariatric surgery, however, the methods developed can be applied to other datasets provided that count/rate data are the focus.

\begin{sloppypar}In the case study here, relatively simple models, such as the intercept-only zero-truncated Poisson, turned out to be fitting well. In other cases, this might be quite different.
 Person-times for each study are included as an exposure variable to allow for more accurate modelling and estimates, leading to the implication that the rate of completed suicide remains constant over time. A sensitivity analysis conducted in the Supplementary Material showed that this assumption is reasonable and that the original model chosen in this paper is preferred.
 Future work relating to this paper could include exploring other methods for population size estimation, such as both Chao's and Zelterman's estimators \citep{bohning2008estimating}, which also allow for the inclusion of covariates for population size estimation, but are more robust in their distributional assumptions. This would enable alternative population sizes to be found, with the possibility for allowing covariate effects and further uncertainty quantification through application of the bootstrap algorithm. Additional future work could include exploring alternative applications of the methods discussed. Whilst the intercept-only zero-truncated Poisson model is preferred for the suicide data, this may not be the case for another meta-analytic dataset. This could also lead to the opportunity of expanding these methods to account for possible one-inflation, as whilst there are many singletons in this data, analysis in the Supplementary Material shows that there is no one-inflation present in the suicide data.
 Another form of heterogeneity occurs with unmeasured covariates, the unobserved heterogeneity. This form of heterogeneity is typically modelled by means of a random effect approach. The negative-binomial distribution can be viewed as such a random effects model, in which the random effects distribution is modelled via a gamma distribution. However, it was seen that there was no gain in working with the negative-binomial instead of the Poisson distribution.    A further alternative could follow to model unobserved heterogeneity by non-parametric mixture models to leave the random effects distribution unspecified \citep{norris1996nonparametric}.  In the case study here, a two-component discrete mixture model has a BIC-value of 57.3 whereas the homogeneous  Poisson model shows a BIC-value of 50.7, indicating again the good fit of the Poisson model and no evidence of residual heterogeneity and therefore no gain in this case in working with an alternative model to the Poisson.\end{sloppypar}

\bibliographystyle{bibstyle}

\end{document}